\documentclass[a4paper,11pt,tikz]{article}
\pdfoutput=1
\usepackage{jheppub}
\usepackage{amsmath,amssymb,bm,graphicx,bbold,epsf,colordvi}
\usepackage{lipsum}
\usepackage{makecell}
\usepackage{soul}
\usepackage{braket}
\usepackage{bm}
\usepackage{appendix}
\allowdisplaybreaks 
\usepackage{color}
\usepackage[dvipsnames]{xcolor}

\usepackage{mathrsfs}
\usepackage{slashed}
\usepackage{graphicx}
\usepackage[dvipsnames]{xcolor}
\usepackage{amsmath}
\allowdisplaybreaks
\usepackage{amssymb}

\usepackage{slashed}
\usepackage[utf8]{inputenc}
\usepackage{hhline,multirow,tabularx}
\usepackage{bm}
\usepackage{hyperref}
\usepackage[export]{adjustbox}
\usepackage{mdframed}
\usepackage{comment}
\usepackage{natbib}
\def\md{\mathrm{d}}

\def\bea#1\eea{\begin{align}#1\end{align}}

\usepackage{lipsum}

\makeatletter
\def\@fpheader{~}
\makeatother

\title{Revisiting Azimuthal Angular Asymmetries in Diffractive Di-jet Production}

\author[a,b,c]{Ding Yu Shao}

\author[d]{, Yu Shi}

\author[a]{, Cheng Zhang}

\author[d]{, Jian Zhou}

\author[d]{and Ya-jin Zhou}

\affiliation[a]{Department of Physics and Center for Field Theory and Particle Physics, Fudan University, Shanghai 200438, China}

\affiliation[b]{Key Laboratory of Nuclear Physics and Ion-beam Application (MOE), Fudan University, Shanghai 200438, China}

\affiliation[c]{Shanghai Research Center for Theoretical Nuclear Physics, NSFC and Fudan University, Shanghai 200438, China}

\affiliation[d]{\normalsize\it Key Laboratory of Particle Physics and
Particle Irradiation (MOE), Institute of Frontier and
Interdisciplinary Science, Shandong University, Qingdao, China}

\emailAdd{dingyu.shao@cern.ch}
\emailAdd{yu.shi@sdu.edu.cn}
\emailAdd{chengzhang\_phy@fudan.edu.cn}
\emailAdd{jzhou@sdu.edu.cn}
\emailAdd{zhouyj@sdu.edu.cn}

\abstract{
We explore the impact of initial state soft gluon radiations on the azimuthal angle asymmetries in photo-production of hard di-jet via coherent diffraction in ultraperipheral heavy ion collisions, as well as in electron-proton ($ep$) and electron-nucleus ($eA$) collisions. The primary production mechanism is identified as the diffractive production of two hard jets, accompanied by a collinear gluon emission along the beam direction. In contrast, the diffractive exclusive di-jet production, where the initial state radiation is absent, is suppressed due to color transparency. Our analysis shows that azimuthal asymmetries, traditionally attributed to final state gluon emissions, are reduced by the presence of initial state radiations. The sensitivity of azimuthal asymmetries to both initial and final state radiations suggests that they could provide novel insights into the mechanisms of di-jet production in diffractive processes.
}

\begin{document}
\maketitle

\section{Introduction}\label{Introduction}

Diffractive di-jet production in hard scattering processes has attracted considerable attention in recent years~\cite{Braun:2005rg,Hatta:2016dxp,Altinoluk:2015dpi,Zhou:2016rnt,Hagiwara:2017fye,Mantysaari:2019csc,Hatta:2019ixj,Mantysaari:2019hkq,Guzey:2019dpp,Guzey:2020ehb,Guzey:2020gkk,Iancu:2021rup,Boer:2021upt,Hatta:2022lzj,Zhang:2022tee,Iancu:2022lcw,Frankfurt:2022jns,Iancu:2023lel,Rodriguez-Aguilar:2023ihz}, as the rich information on nucleon internal structure can be extracted via these processes. The primary theoretical emphasis lies in the correlation limit, where the two measured jets are relatively hard and nearly back-to-back in the transverse plane, enabling a factorization of the interesting parton distributions from the comparatively hard physics underlying dijet production. At large or moderate $x$,  diffractive exclusive dijet production in electron-proton ($ep$) collisions emerges as a sensitive probe of the generalized  parton distribution (GPD) of gluon in the nucleon~\cite{Ji:2003ak, Belitsky:2003nz,Braun:2002en,Braun:2005rg}. In this aspect, the dijet production with large invariant mass is complementary to vector meson production and may offer some advantages as higher twist effects are suppressed. On the other hand, in the gluon saturation regime at small $x$, it was suggested~\cite{Hatta:2016dxp,Altinoluk:2015dpi,Zhou:2016rnt,Hagiwara:2017fye,Mantysaari:2019csc,Mantysaari:2019hkq,Boer:2021upt} that diffractive exclusive di-jet production in $ep/eA$ collisions may give direct access to the gluon Wigner distribution which encodes the complete information about how gluons are distributed both in position and momentum spaces. Additionally, the possibility of extracting the canonical gluon orbital angular momentum from diffractive di-jet production in polarized $ep$ collisions has been proposed in Refs.~\cite{Ji:2016jgn,Hatta:2016aoc,Bhattacharya:2022vvo,Bhattacharya:2018lgm,Boussarie:2018zwg}, highlighting its potential in advancing our understanding of nucleon inner structure.

The Quantum Chromodynamics (QCD) analysis of diffractive di-jet production poses a fascinating challenge~\cite{Collins:1997sr}. In cases of diffractive tri-jet production characterized by an asymmetric configuration, where a semi-hard gluon is emitted towards the target direction and remains undetected, the experimental signature of this process becomes indistinguishable from that of exclusive di-jet production. Recent studies~\cite{Iancu:2021rup,Iancu:2022lcw} have shown that the cross section for coherent tri-jet photo-production significantly surpasses that of exclusive di-jet production, despite being formally considered as a higher-order correction. The ZEUS Collaboration's  measurements  on diffractive photo-production and electron-production of dijets lend substantial support to this finding~\cite{ZEUS:2015sns}. At LHC energies, the exclusive di-jet cross section is expected to be two to three orders of magnitude lower than the semi-inclusive cross section associated with the (2+1)-jet channel in Ultra-Peripheral Collisions (UPCs)~\cite{Iancu:2023lel}. Given that the quark-antiquark pair in the tri-jet configuration remains in a color octet state and is closely spaced in the transverse coordinate space, this setup can effectively be viewed as the elastic scattering of a gluon-gluon dipole. This configuration allows for strong scattering between the $q\bar{q}g$ system and the nucleus target, circumventing the higher twist suppression typically seen in exclusive dijet production due to color transparency. Consequently, this process has been argued to offer greater sensitivity to gluon saturation effects~\cite{Iancu:2021rup,Iancu:2022lcw,Iancu:2023lel}. The tri-jet cross-section, as formulated within the Color Glass Condensate (CGC) formalism~\cite{Mueller:1993rr,Mueller:1999wm,McLerran:1993ni,McLerran:1993ka,McLerran:1994vd}, can be further factorized into a convolution of the hard part and the diffractive transverse momentum dependent (DTMD) parton distribution function, which can be calculated in terms of the dipole amplitude~\cite{Hebecker:1997gp,Buchmuller:1998jv,Golec-Biernat:1999qor,Hautmann:1999ui,Hautmann:2000pw,Golec-Biernat:2001gyl,Hatta:2022lzj,Salajegheh:2023jgi}.

The crucial observable for exploring gluon tomography in the nucleon/nucleus is the transverse momentum imbalance between the two hard jets $\bm q_\perp=\bm k_{1\perp}+\bm k_{2\perp}$ where $\bm k_{1\perp}$ and $\bm k_{2\perp}$ denote the individual jets' transverse momenta, respectively.  In particular, the nontrivial azimuthal modulations of the cross section encode the novel partonic structure of the target. For instance, the elliptic gluon Wigner distribution generates a $\cos(2\phi)$  asymmetry, where $\phi$ is the angle between $\bm q_\perp$ and $\bm P_\perp=(\bm k_{1\perp}-\bm k_{2\perp})/2$~\cite{Hatta:2016dxp,Altinoluk:2015dpi,Zhou:2016rnt,Boussarie:2018zwg,Hatta:2017cte,Mantysaari:2019csc,Mantysaari:2020lhf,Dumitru:2021mab,Hagiwara:2021xkf}.  Two primary factors contribute to the deviations from the exact back-to-back configuration in diffractive exclusive di-jet production: transverse momentum carried by the pomeron and final state gluon radiations from the jets.  In the case of tri-jet production, two additional contributions give rise to the momentum imbalance: intrinsic gluon transverse momentum dependent (TMD) distribution inside the exchanged pomeron and initial state soft gluon radiations. 
 Extracting information on nucleon structure requires an accurate theoretical account of the pure perturbative QCD origin of the transverse momentum imbalance.   A primary example is that final state gluon radiations can produce the same  $\cos(2\phi) $ azimuthal asymmetry as the elliptic gluon distribution.  Such non-trivial azimuthal modulation arises because the emitted soft gluons tend to be aligned with jet directions.  Refs.~\cite{Hatta:2020bgy, Hatta:2021jcd} carry out a comprehensive study of the impact of final state radiations on the azimuthal angle correlations.  The implications of these contributions on probing saturation physics and studying UPC observables have been addressed in Refs.~\cite{Mantysaari:2017slo, Xing:2020hwh,Hagiwara:2020juc,Hagiwara:2021xkf, Brandenburg:2022jgr,Shao:2022stc,Shao:2023zge}.

In this paper, we study how $q_\perp$ distribution and azimuthal asymmetries are affected by the initial state radiations in diffractive tri-jet production, although in practice, only two hard jets are measured in experiments. Following recent developments~\cite{Iancu:2021rup,Iancu:2022lcw,Iancu:2023lel,Hatta:2022lzj},  we compute the diffractive tri-jet production using the diffractive TMD factorization approach.  Within this formalism, the gluon DTMD distribution naturally emerges as one of the basic ingredients in the factorized cross section formula. As such, initial state soft gluon radiation contributions can be resummed to all orders and incorporated into the scale-dependent gluon DTMD by solving the standard Collins-Soper equation and the renormalization group equation. 

Our findings show that the $q_\perp$ distribution gets broadened due to the initial state radiations, aligning with theoretical expectations. Additionally, gluons are emitted from incoming partons in an axially symmetrical fashion, which tends to smear out azimuthal asymmetries induced by final-state gluon radiations.  The numerical estimations of the azimuthal asymmetries for diffractive di-jet production in UPCs were carried out and tested against the CMS measurements. Predictions for the asymmetries in diffractive photo-production of di-jets in electron-nucleus ($eA$) collisions at the Electron-Ion Collider (EIC) were also formulated. We propose that the azimuthal asymmetries explored herein serve as sensitive probes to distinguish between different production mechanisms in hard diffractive scattering processes. Furthermore, the precise determination of perturbative QCD background is also crucial to obtain a multi-dimensional image of gluonic matter in the target from the diffractive di-jet production. 

The paper is structured as follows. In Section \ref{sec:CGC_LO}, we briefly review the CGC calculation of diffractive tri-jet photo-production at the Born level. We then discuss the factorization scheme employed in our calculation and the associated resummation formula in Section \ref{sec:fac_res}.  The numerical results are presented in Section \ref{sec:pheno}.  Finally, the paper concludes with a summary in Section \ref{sec:conclusion}.

\section{CGC calculation of semi-inclusive diffractive di-jet photo-production}\label{sec:CGC_LO}

The CGC calculation of diffractive di-jet photo-production, accompanied by a semi-hard gluon emission, has been formulated in Refs.~\cite{Iancu:2021rup,Iancu:2022lcw,Iancu:2023lel}. Let us briefly review this calculation by first specifying the relevant kinematics,
\begin{eqnarray}
\gamma(x_\gamma p)+A \rightarrow q(k_1)+\bar q(k_2)+g(l)+A,
\end{eqnarray}
where $\gamma$ represents a quasi-real photon, and $A$ denotes the nuclear target. In $ep$ collisions, the electron emits this quasi-real photon, with $x_\gamma$ indicating the fraction of the electron's momentum transferred to the photon. In UPCs, this quasi-real photon is alternatively emitted from one of the colliding nuclei, with the other nucleus serving as the target. Intriguingly, in the observed final state, it is indistinguishable whether the photon originated from the projectile or the target nucleus, thereby manifesting a double-slit interference phenomenon at the Fermi scale~\cite{Klein:1999gv,Zha:2018jin,Xing:2020hwh,Mantysaari:2023prg}. However, this interference effect is predominantly noticeable at very low pair transverse momentum, approximately $30$ MeV (comparable to the inverse of the nuclear radius), and can be disregarded for the present analysis focused on the semi-hard region.

In the process under consideration, the quasi-real photon initially decays into a quark-antiquark pair. Subsequently, one of these particles, either the quark or the antiquark, emits a gluon. Following this emission, the three partons – the quark, antiquark, and gluon – undergo elastic scattering off the nuclear target. To quantitatively describe this process, we define the longitudinal momentum fractions of the quasi-real photon carried by the quark, antiquark, and gluon, respectively. These fractions are denoted as,
\begin{align}
    z_1=\frac{k_1^+}{x_\gamma P^+},~~z_2=\frac{k_2^+}{x_\gamma P^+},~~{\rm and}~~ z_3=\frac{l^+}{x_\gamma P^+},
\end{align}
where  the incoming electron or nucleus is assumed to move along the $z$ direction and carries the momentum $ P^+ p^\mu$ with $ p^\mu\equiv1/\sqrt{2}(1,0, 0_\perp)$ defined in the light-cone coordinate. Obviously, one has $z_1+z_2 +z_3=1$ which reflects momentum conservation along the light-cone direction.  Furthermore, the longitudinal momentum fractions of the nucleon carried by the quark, antiquark, and gluon are represented as
\begin{align}
    x_1=\frac{k_1^-}{\bar P^-},~~ x_{2} =\frac{k_2^-}{\bar P^-}~~ {\rm and}~~ x_3=\frac{l^-}{\bar P^-}.
\end{align}
where $\bar P^-$ is the dominant component of light cone momentum carried by the nucleus target. Additionally, the longitudinal fractions attributed to the pomeron and the hard di-jet system are expressed by $x_{\mathbb{P}}=x_1+x_{2}+x_3$ and $x_{q \bar q}=x_1+x_{2}$, respectively. In this context, $x_{\mathbb{P}}$ should be sufficiently small to ensure coherent scattering, allowing the target to remain intact after collision.

As previously mentioned, we focus on asymmetric 3-jet configurations characterized by significantly harder quark and antiquark jets compared to the third gluon jet. Specifically, the kinematic conditions of interest are defined by the relations $l_\perp \sim q_\perp \ll k_{1\perp} \approx k_{2\perp}$, and $z_3 \ll z_1 \approx 1-z_{2}$. This configuration facilitates a clear factorization between the hard contribution from the quark and antiquark jets and the semi-hard contribution from the gluon jet. Beyond these kinematic regions, the cross section is significantly suppressed.

To intuitively understand the process, consider the following outlined steps. Initially, the target emits a pomeron, which then emits two gluons in a color singlet configuration. One gluon is emitted into the $s$ channel, carrying a momentum fraction $x_3$ and transverse momentum $\bm l_\perp$, while the second gluon, involved in the $t$ channel, carries a momentum fraction $x_{q \bar q}$ and transverse momentum $\bm q_\perp$. The interaction between the quasi-real photon and the $t$-channel gluon leads to the production of two hard jets. Concurrently, the $s$-channel gluon, not directly observed, is recognized as the source of the third, softer gluon jet. This factorization framework enables the formulation of the Born cross section for semi-inclusive diffractive dijet production, as detailed in references~\cite{Iancu:2021rup,Iancu:2022lcw,Iancu:2023lel}. Explicitly, it is expressed as
\begin{eqnarray}\label{eq:borncs}
\frac{\md \sigma}{\md y_1 \, \md y_2 \, \md^2 \bm P_\perp \md^2 \bm q_\perp }=\sigma_0 x_\gamma f_\gamma(x_\gamma)  \int \frac{\md x_{\mathbb{P}} }{x_{\mathbb{P}} } x_g G_{\mathbb{P}}(x_g,x_{\mathbb{P}} ,q_\perp) ,
\end{eqnarray}
where the rapidity of the third gluon jet has been integrated out.  Here, $y_1$ and $y_2$ are the quark and antiquark rapidities, respectively. The hard fact $\sigma_0$ describes the partonic scattering process $\gamma+g \rightarrow q+\bar q $.  At the tree level, it is given by
\begin{eqnarray}
  \sigma_0=\sum_f \alpha_{e} \,\alpha_s \, e_f^2 \,  z_1 (1-z_1) \left [ z_1^2+(1-z_1)^2  \right 
 ]\frac{1}{P_\perp^4},
\end{eqnarray}
where $\alpha_e$ is the fine structure constant, $\alpha_s$ is the strong coupling constant and $e_f$ represents the fractional charge of the quark flavor $f$ under consideration.

In Eq.~\eqref{eq:borncs}, the function $f_\gamma(x_\gamma)$ represents the photon distribution function within an electron or a large nucleus. Specifically, in the UPC case, the photon flux generated by one of the incoming nuclei has to be integrated over the impact parameter range $[2R_A, \infty)$ where $R_A $ is the radius of the nucleus.  A classical electrodynamics calculation of the collinear photon distribution yields
~\cite{Bertulani:1987tz,Bertulani:2005ru, Baltz:2007kq},
\begin{eqnarray}
x_\gamma f_\gamma(x_\gamma)= \frac{2 Z^2 \alpha_{e}}{\pi} \left [  \zeta K_0(\zeta) K_1(\zeta) -\frac{\zeta^2}{2} \left ( K_1^2(\zeta)-K_0^2(\zeta) \right ) \right ],
\end{eqnarray}
where $\zeta\equiv 2x_\gamma M_p R_A $ and $M_p$ is the proton mass. Besides, $Z$ is the nuclear charge number, and $K_0(\zeta)$ and $K_1(\zeta)$ are modified Bessel functions of the second kind. Photon energy can be expressed in terms of rapidities and jet transverse momentum as $x_\gamma =\frac{P_\perp}{\sqrt{s}}(e^{y_1}+e^{y_2})$ with $\sqrt{s}$ being the center of mass energy per nucleon pair. In the case of $ep $ collisions, the photon PDF within the electron is computed at leading order in QED as follows:
\begin{eqnarray}
   f_\gamma(x_\gamma,\mu^2)=   \frac{\alpha_e}{2\pi}  \frac{1+(1-x_\gamma)^2}{x_\gamma} \ln \frac{ \mu^2}{x_\gamma^2 m_e^2}, 
  \end{eqnarray}
where $m_e$ is the electron mass, and $\mu$ is the factorization scale which will be specified later.  Although a leading-order calculation, this provides sufficient accuracy for our purposes. For a more comprehensive analysis, one could implement DGLAP evolution to refine the photon PDF at any specified factorization scale. This approach, widely recognized in the literature (see, for example, Ref.~\cite{Liu:2020rvc}), is utilized for scaling distribution functions across various energy scales.

The gluon diffractive TMD $G_{\mathbb{P}}(x,x_{\mathbb{P}} ,q_\perp)$ offers a clear probability interpretation, describing the likelihood of finding a gluon with a momentum fraction of pomeron $x=x_{q \bar q}/x_\mathbb{P}$ inside a pomeron that carries momentum fraction $x_{\mathbb{P}}$ of the nucleon. These variables are determined by the external kinematics: $x_g=x_{q\bar q} = \frac{P_\perp}{\sqrt{s} } \big(  e^{-y_1} + e^{-y_2} \big)$ and $x_{\mathbb{P}} = x_{q\bar q}  + \frac{q_\perp }{\sqrt{s} }e^{-y_3}$. The transverse momentum transfer to the dijet system via the exchanged gluon is represented by $q_\perp$. Within the CGC formalism, the gluon distribution of the pomeron is related to the gluon-gluon dipole scattering amplitude, as given by~\cite{Iancu:2021rup,Iancu:2022lcw,Iancu:2023lel},
\begin{eqnarray}
 xG_{\mathbb{P}}(x,x_{\mathbb{P}} ,q_\perp)\!   =\! \frac{S_\perp (N_c^2-1)}{8\pi^4 (1\! -x) }  \left [  \frac{x q_\perp^2}{1\!-\!x} \int \! r_\perp \md r_\perp J_2(q_\perp r_\perp) K_2 \left (\! \sqrt{\frac{x q_\perp^2 r_\perp^2  }{1-x}} \right ) {\cal T}_g(x_{\mathbb{P}}, r_\perp) \right]^2, \label{pgluonTMD}
\end{eqnarray}
where $S_\perp$ is the transverse area of the nucleus. The dipole amplitude ${\cal T}_g(x_{\mathbb{P}}, r_\perp)$ can be either computed using the McLerran-Venugopalan (MV) model or parametrized by the Golec-Biernat-Wüsthoff (GBW) model for energies that are not excessively high.  In the dilute limit, the gluon distribution of pomeron scales as $1/q_\perp^4$, which implies that the typical transverse momentum of the third jet is of the order of the saturation scale. The rapidity of the third jet is so large that it is unlikely to be probed by detectors. 

Our primary focus of this work is the distribution of the transverse momentum imbalance of the two hard jets and the azimuthal angle correlation. By neglecting the transverse momentum carried by the pomeron, the two hard jets acquire the transverse momentum imbalance entirely from the recoil effect, which implies $\bm q_\perp=-\bm l_\perp$ at the Born level. However, beyond the tree level, the Collins-Soper type evolution, which accounts for initial state and final state soft gluon radiations, eventually controls the dijet imbalance. On the one hand, the azimuthal asymmetry arises from final state gluon radiation which tends to align with the directions of hard jets. On the other hand, these asymmetries are attenuated by initial-state radiations. These effects involve contributions that are enhanced by large double or single logarithm terms, necessitating all-order resummation. We will discuss the factorization scheme and the associated resummation procedure in detail in the next section. 

\section{Factorizaton and resummation formula}\label{sec:fac_res}

\begin{figure}
    \centering
    \includegraphics[width=1\linewidth]{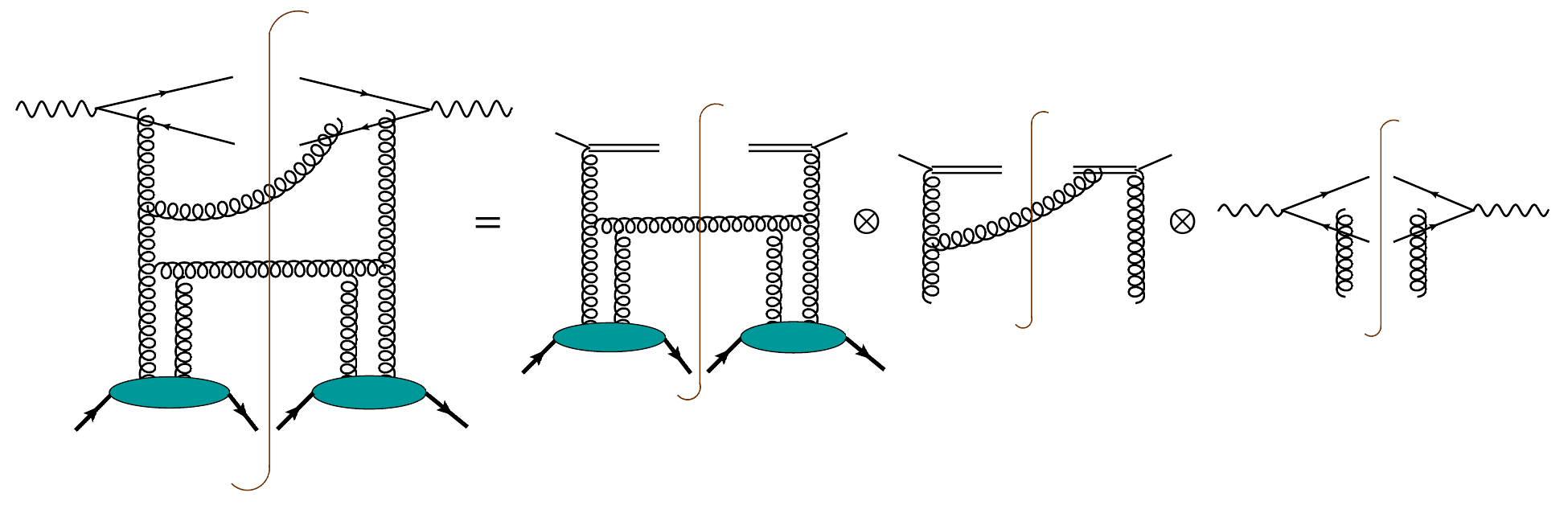}
    \caption{An illustration of the factorization scheme. The double line represents the gauge link. }
    \label{fig:enter-label}
\end{figure}

In the back-to-back region, characterized by the conditions $P_\perp \gg q_\perp$, the differential cross section can be factorized within the soft-collinear effective theory (SCET)~\cite{Bauer:2000ew,Bauer:2000yr,Bauer:2001ct,Bauer:2001yt,Bauer:2002nz} framework.  As illustrated in Fig. \ref{fig:enter-label}, this factorization is expressed as:
\begin{align}\label{eq:fac-mom}
    \frac{\md \sigma}{\md y_1 \, \md y_2 \, \md^2 \bm P_\perp \, \md^2 \bm q_\perp }=&\,\sigma_0  x_\gamma f_\gamma(x_\gamma) H_{\gamma^* g}(P_\perp,R,\mu) \int \md^2 \bm k_\perp \md^2 \bm \lambda_\perp \delta^{(2)}(\bm q_\perp - \bm k_\perp - \bm \lambda_\perp) 
      \notag \\
     &\times  S(\bm \lambda_\perp,R,\mu) \int \frac{\md x_{\mathbb{P}}}{x_{\mathbb{P}}} x_g G_{\mathbb{P}}^{\rm unsub}(x_g, x_{\mathbb{P}}, k_\perp, \mu).
\end{align}
In this formulation, we neglect the transverse momentum dependence of the incoming photon due to QED radiations. The Dirac delta function $\delta^{(2)}(\cdots)$ enforces transverse momentum conservation. To account for higher order corrections, we introduce the scale-dependent pomeron gluon TMD function $x G_{\mathbb{P}}^{\rm unsub}(x, x_{\mathbb{P}},  k_\perp, \mu)$, which describes the transverse momentum of the gluon relative to the pomeron at the given factorization scale $\mu$. Furthermore, $H_{\gamma^* g}$ is the hard function that describes perturbative corrections for the hard scattering process in dijet production due to photon and gluon fusion, and $H_{\gamma^* g}=1$ at the leading order. The soft function $S(\bm \lambda_\perp,R,\mu)$ captures the soft radiations from the incoming gluon and the final-state jets. The dependence on the jet radius $R$ in both hard and soft functions reflects the jet definition used in this context.

To facilitate carrying out the resummation of the TMD observable, it is convenient to convert the above factorization formula to the transverse coordinate space. After applying a Fourier transform to Eq. \eqref{eq:fac-mom}, we obtain the factorized formula in the coordinate space of $b_\perp$ as follows:
\begin{align}
    \frac{\md \sigma}{\md y_1 \, \md y_2 \, \md^2 \bm P_\perp \, \md^2 \bm q_\perp } =&\,\sigma_0  x_\gamma f_\gamma(x_\gamma) H_{\gamma^* g}(P_\perp,R,\mu) \int \frac{\md^2 \bm b_\perp}{(2\pi)^2} 
     e^{i \bm q_\perp \cdot \bm b_\perp}  \tilde S(\bm b_\perp,R,\mu) \notag \\
     &\times  \int \md^2 \bm k_\perp e^{- i \bm b_\perp \cdot \bm k_\perp}  \int \frac{\md x_{\mathbb{P}}}{x_{\mathbb{P}}} x_g G_{\mathbb{P}}^{\rm unsub}(x_g, x_{\mathbb{P}}, k_\perp, \mu).
\end{align}
Similar to conventional TMDs, the naive definition of the pomeron gluon TMD function exhibits rapidity singularities at higher orders. These singularities are removed by subtracting the TMD soft function as suggested in the Collins-11 scheme~\cite{Collins:2011zzd}. This subtraction allows for defining ingredients independent of rapidity divergence, as in
\begin{align}\label{eq:Pgluon_redef}
    G_{\mathbb{P}}^{\rm unsub}(x_g, x_{\mathbb{P}}, k_\perp, \mu) \tilde S(\bm b_\perp,R,\mu) = G_{\mathbb{P}}(x_g, x_{\mathbb{P}}, k_\perp, \mu,\zeta) \tilde S^{\rm rem}(\bm b_\perp,R,\mu),
\end{align}
with
\begin{align}
    \tilde S^{\rm rem}(\bm b_\perp,R,\mu) \equiv \frac{\tilde S(\bm b_\perp,R,\mu)}{\sqrt{\tilde S_g(b_\perp,\mu)}}. 
\end{align}
In this context, $\tilde S_g$ is the standard TMD soft function used in Semi-Inclusive Deep Inelastic Scattering (SIDIS) processes. It differs from the usual function by having the Wilson line in the adjoint representation instead of the fundamental representation. Besides, $\tilde S^{\rm rem}$ represents the remaining soft factor. The large logarithms involving the Collins-Soper scale, denoted as $\zeta$,  can be resummed to all orders using the Collins-Soper equation in the perturbative region~\cite{Collins:2011zzd,Boussarie:2023izj}. Non-perturbative effects are parametrized by introducing a non-perturbative Sukakov factor, which will be specified later.

We refactorize the pomeron gluon TMD function as the matching coefficients and the integrated pomeron gluon function as follows:
\begin{align}
    G_{\mathbb{P}}(x_g, x_{\mathbb{P}}, k_\perp, \mu,\zeta) = \int_{x_g}^1 \frac{\md z}{z} I_{g\leftarrow g}(z,k_\perp,\mu,\zeta) G_{\mathbb{P}}(x_g/z, x_{\mathbb{P}}, \mu)+ G_{\mathbb{P}}(x_g,x_{\mathbb{P}} ,k_\perp),
\end{align}
where $I_{g\leftarrow g}$ represents the gluon-to-gluon splitting process. At the leading order,  the matching coefficient is simply given by $I_{g\leftarrow g}=\delta(1-z)\delta^{(2)}(\bm k_\perp)$. At a low initial scale $\mu_0$ one can reconstruct integrated gluon diffractive PDF $G_{\mathbb{P}}(x_g/z, x_{\mathbb{P}}, \mu_0)$   using the relation $G_{\mathbb{P}}(x_g, x_{\mathbb{P}}, \mu_0)=\int \md^2 \bm k_\perp G_{\mathbb{P}}(x_g,x_{\mathbb{P}} ,k_\perp)\theta(\mu_0-k_\perp)$ where $G_{\mathbb{P}}(x_g,x_{\mathbb{P}} ,k_\perp)$ is given in Eq.~\eqref{pgluonTMD}.   We extract the double and single leading logarithm terms from the higher-order matching coefficient $I_{g\leftarrow g}(z,k_\perp,\mu,\zeta)$ and resum them into the Sudakov factor. Among the residual terms in higher-order matching coefficient, those enhanced by the ``collinear divergence" are recognized as the conventional gluon splitting kernel that drives the DGLAP evolution of pomeron gluon PDF. One should notice that the collinear divergence in its true sense is screened by the saturation effect. Apart from this contribution, the leading order term from $ G_{\mathbb{P}}(x,x_{\mathbb{P}} ,k_\perp)$ acts as an additional static source term in the modified DGLAP equation derived in Ref.~\cite{Iancu:2023lel}. However, in this study, our primary interest lies in analyzing the azimuthal angular asymmetry distribution. The impact of the DGLAP evolution becomes significantly reduced once we take the ratio for the azimuthal asymmetry or perform the normalization of the $q_\perp$ distribution. Therefore, for the current purpose, we choose to neglect the DGLAP evolution effect in our numerical calculations.

After evolving the hard function from the hard scale to $q_\perp$, we derive the all-order resummation formula as follows: 
\begin{align}\label{eq:resum_pert}
    \frac{\md \sigma}{\md y_1 \, \md y_2 \, \md^2 \bm P_\perp \, \md^2 \bm q_\perp }=&\,\sigma_0  x_\gamma f_\gamma(x_\gamma) \int \frac{\md^2 \bm b_\perp}{(2\pi)^2} 
    e^{i \bm q_\perp \cdot \bm b_\perp} e^{-\text{Sud}_{\rm pert}(b_\perp)} \tilde S^{\rm rem}(\bm b_\perp,\mu_b) \notag \\
     &\times   \int \md^2 \bm k_\perp e^{- i \bm b_\perp \cdot \bm k_\perp} \int \frac{\md x_{\mathbb{P}}}{x_{\mathbb{P}}} x_g G_{\mathbb{P}}(x_g, x_{\mathbb{P}}, k_\perp),
\end{align}
where the perturbative Sudakov factor is determined by the evolution function of the hard function, expressed as
\begin{align}
    \text{Sud}_{\rm pert}(b_\perp)&=\int_{\mu_b}^{P_\perp} \frac{d \mu}{\mu} \big[\Gamma_{\rm virt.}(\alpha_s) + 2 \, \Gamma_{\rm jet}(\alpha_s)\big],
\end{align}
choosing the hard scale as $P_\perp$. The anomalous dimensions are defined as
\begin{align}
    \Gamma_{\rm virt.}(\alpha_s) &= - C_A \frac{\alpha_s}{\pi} \ln \frac{\mu^2}{P_\perp^2} - 2\, C_F \frac{\alpha_s}{\pi} \ln \frac{\mu^2}{M^2} -3\, C_F \frac{\alpha_s}{\pi}  -2\,C_A \beta_0 \frac{\alpha_s}{\pi}, \\
    \Gamma_{\rm jet}(\alpha_s)&= -  C_F \frac{\alpha_s}{\pi}\ln \frac{P_{\perp}^2 R^2}{\mu^2}+ 3\, C_F \frac{\alpha_s}{2\pi},
\end{align}
where $\beta_0=11/12-N_f/18$, and $C_A$ and $C_F$ are the Casimir operators for the adjoint and fundamental representations, respectively. The term $\Gamma_{\rm virt.}$ denotes the contribution from virtual corrections in the $\gamma^* g \to q\bar q$ process, while $\Gamma_{\rm jet}$ accounts for contributions from energetic radiations inside the jet with radius $R$. Consequently, the perturbative Sudakov factor is obtained as
\begin{align}
    \text{Sud}_{\rm pert}(b_\perp) = C_A \int_{\mu_b}^{P_\perp} \frac{\md \mu}{\mu} \frac{\alpha_s}{\pi} \left(\ln\frac{P_\perp^2}{\mu^2} + \frac{2\,C_F}{C_A} \ln \frac{M^2}{P_\perp^2 R^2} - 2\, \beta_0\right).
\end{align}

Since our interest lies in the azimuthal asymmetry distribution, significantly influenced by soft radiation in the perturbative region, we keep the azimuthal dependent part in the soft function $\tilde S^{\rm rem}$ in the resummation formula \eqref{eq:resum_pert}. The next-to-leading order (NLO) soft function is expressed as
\begin{align}\label{eq:nlo_softfun}
   \tilde S^{\rm NLO}(\bm b_\perp,R,\mu) = \frac{C_A}{2} \, \omega_{gq} + \frac{C_A}{2} \, \omega_{g\bar q} + \left(C_F-\frac{C_A}{2}\right) \omega_{q\bar q}.
\end{align}
Here $\omega_{ij}$ represents the soft gluon phase space integration, defined as
\begin{align}
    \omega_{ij}= &  \frac{\alpha_s \mu^{2 \epsilon} \pi^\epsilon e^{\gamma_E \epsilon}}{\pi^2} \int \md^d k \, \delta \!\left(k^2\right) \theta\!\left(k^0\right) \frac{n_i \cdot n_j}{n_i \cdot k \, k \cdot n_j}\left(\frac{\nu}{2\, k^0}\right)^\eta \theta(\Delta R_q-R) \theta(\Delta R_{\bar q}-R)  e^{-i \bm{k}_{\perp} \cdot \bm{b}_\perp},
\end{align}
where $n_i$ refer to the directions of initial and final-state partons, with $i,j=g,q,\bar q$. The term $\Delta R_{q,(\bar q)}$ denotes the distance between the jet and the soft emission in the rapidity and azimuthal angle plane, defined as $\Delta R_i \equiv \sqrt{\Delta \phi_i^2 + \Delta y_i^2}$. The condition $\theta(\Delta R_i-R)$ ensures that the soft gluon with momentum $k$ is emitted outside the jet boundary, as soft radiations within jets contribute only in the region of $q_\perp=0$. The necessity of the rapidity regulator $\eta$ in $\omega_{gq}$ and $\omega_{g\bar q}$ arises due to rapidity divergence, which is not addressed by dimensional regularization. This rapidity regulator and its associated scale dependence are later removed, as defined in Eq. \eqref{eq:Pgluon_redef}. Detailed calculations are presented in the appendix \ref{app:softfun}, and in the narrow cone approximation, the azimuthal angle-dependent one-loop soft factor is
\begin{align}\label{eq:s_NLO_phidep}
    \tilde S^{\rm NLO}_{\phi_b{\rm -dep}}(\bm b_\perp,R,\mu_b)= &\, \frac{\alpha_s(\mu_b)}{\pi} \Bigg\{ C_F\left( - \ln\frac{4}{R^2}\ln c_\phi^2 - \frac{1}{2}\ln^2c_\phi^2 \right) + \frac{1}{2N_c} \Bigg[ - \frac{1}{2} \ln^2 c_\phi^2 +  \ln c_\phi^2 \ln x  \notag \\
    & -  \ln c_\phi^2 \ln\left(1 - \frac{x}{c_\phi^2}\right) +  \ln x \log\left(1 - \frac{x}{c_\phi^2}\right) +  \, \text{Li}_2\left(\frac{x}{c_\phi^2}\right)
     \Bigg]\Bigg\},
\end{align}
with $\mu_b = 2\,e^{-\gamma_E}/b_\perp$, $x=M^2/(4\,P_\perp^2)$ and $c_\phi = \cos\phi_b$, where $\phi_b$ is the azimuthal angle between the impact parameter $\bm b_\perp$ and $\bm P_\perp$. In these calculations, we have ignored the power corrections from the jet radius $R$ and retained only its logarithmically dependent terms. After performing the azimuthal angle projection, we obtain
\begin{align}
    \tilde S^{\rm rem}(\bm b_\perp, \mu_b) = 1-c_2 \frac{2\,\alpha_s(\mu_b) C_F}{\pi} \cos 2\phi_b + \cdots. 
\end{align}
The coefficient $c_2$ reads
\begin{align}
    c_2=\ln \frac{1}{e R^2}-\frac{1}{2\,C_F N_c}\ln(a_2 e), 
\end{align}
with $ a_2\equiv \exp[\Delta y \sinh \Delta y-\cosh \Delta y \ln [2(1+\cosh \Delta y)]]$, which is consistent with those found in Ref. \cite{Hatta:2021jcd}. The underlying physics behind this azimuthal asymmetry is well understood: soft gluons are more likely to be emitted along the jet direction, resulting in an enhanced differential cross section in the region where $\bm q_\perp$  is aligned or anti-aligned with $\bm P_\perp$. The azimuthal dependent cross section eventually can be cast into the following form,
\begin{align}\label{eq:cos2phi}
    \frac{\md \sigma}{\md y_1 \, \md y_2 \, \md^2 \bm P_\perp \, \md^2 \bm q_\perp }=&\,  \sigma_0  x_\gamma f_\gamma(x_\gamma) \int \frac{\md^2 \bm b_\perp}{(2\pi)^2} 
    e^{i \bm q_\perp \cdot \bm b_\perp} e^{-\text{Sud}_{\rm pert}(b_\perp)}\left[1-c_2 \frac{2\,\alpha_s(\mu_b) C_F}{\pi} \cos 2\phi_b \right] \notag \\
     &\times   \int \md^2 \bm k_\perp e^{- i \bm b_\perp \cdot \bm k_\perp} \int \frac{\md x_{\mathbb{P}}}{x_{\mathbb{P}}} x_g G_{\mathbb{P}}(x_g, x_{\mathbb{P}}, k_\perp),
\end{align}
which is the central result of this work. We are now ready to proceed with the numerical estimations. 

\section{Phenomenology studies}\label{sec:pheno}

In this section, we present the numerical results for $q_\perp$ distribution and azimuthal asymmetries for semi-inclusive di-jet production in UPCs at LHC energy and compare them with the CMS measurements. Additionally, we extend our predictions to analogous observables in diffractive photo-production of di-jets at both the EIC and HERA. To begin, we introduce the essential components and parameters required for our numerical estimations.

First of all, in addition to the perturbative Sudakov factor, we introduce the non-perturbative Sudakov factor in the resummation formula to account for non-perturbative effects as $\mu_b$ approaches $\Lambda_{\rm QCD}$. Specifically, in Eq.~\eqref{eq:resum_pert}, we modify the perturbative Sudakov factor, $\text{Sud}_{\rm pert}(b_\perp)$, as follows:
\begin{align}
    \text{Sud}_{\rm pert}(b_\perp) \to \text{Sud}(b_\perp) \equiv 2\,\text{Sud}^{\rm jet}_{\rm NP}(b_\perp) +  \frac{C_A}{C_F} \text{Sud}_{\rm NP}(b_\perp, P_\perp) +\text{Sud}_{\rm pert}(b_\perp^*, P_\perp), 
\end{align} 
where the non-perturbative components of the Sudakov factor are defined as
\begin{equation}
\text{Sud}^{\rm jet}_{\rm NP}(b_\perp)
 = 
g_\Lambda b_\perp^2, \qquad g_\Lambda =0.1\,  {\rm GeV}^2, 
\end{equation}
and
\begin{equation}
\text{Sud}_{\rm NP }(b_\perp, Q)
=0.106 \, b^2_\perp+0.42\,\ln\frac{Q}{Q_0}\ln \frac{b_\perp}{b_*},
\end{equation}
with $Q^2_0=2.4\,$GeV$^2$ \cite{Su:2014wpa,Echevarria:2020hpy}. Besides, $\mu_b$ in the perturbative part of the Sudakov factor and soft factor is replaced by $\mu_{b_*}\equiv 2\,e^{-\gamma_E}/b_{*}$ with $b_*=b_\perp/\sqrt{1+b_\perp^2/b^2_{\rm max}}$ and $b_{\rm max}=1.5\,$GeV$^{-1}$. This formulation uses the $b_*$-prescription to regularize the impact parameter $b_\perp$ in the limit $b_\perp \to \infty$, which corresponds to the infrared region.  If the value of $\mu_ b$ is larger than $P_\perp R$, we set $\mu_ b=P_\perp R$.  In this work, we only consider the leading double logarithm and leading single logarithm contributions, so we need to use the one-loop running coupling $\alpha_s(\mu)$, which is defined as
\begin{eqnarray}
\alpha_s(\mu)= \frac{12\pi}{(33-2N_f)\ln(\mu^2/\Lambda_{\rm QCD}^2)},
\end{eqnarray}
with $\Lambda_{\rm QCD}=0.24\,$GeV. The perturbative part of the Sudakov factor can be further decomposed into two terms,
 \begin{equation}
\text{Sud}_{\rm pert}(b_\perp^*, P_\perp) =\text{Sud}^{ i}(b_\perp^*, P_\perp)+\text{Sud}^{ f}(b_\perp^*, P_\perp),
\end{equation}
where 
\begin{equation}
\text{Sud}^{ i}(b_\perp^*, P_\perp)\equiv 
\int_{{\mu_{b*}}}^{P_\perp} \frac{\md \mu }{\mu} \frac{\alpha_s(\mu) C_A}{\pi} \left [   \ln\left(\frac{P_\perp^2}{\mu^2}\right)-2 \,\beta_0  \right ],
\end{equation}
and 
\begin{equation}
\text{Sud}^{ f}(b_\perp^*, P_\perp)\equiv
\int_{\mu_{b*}}^{P_\perp} \frac{\md \mu}{\mu} \frac{\alpha_s(\mu)C_F}{\pi} 2 \ln \frac{M^2}{P_\perp^2 R^2},
\end{equation}
respectively. To fully expose the impact of initial state gluon radiation effects, which were not considered in the previous analysis~\cite{Hatta:2021jcd}, on the $q_\perp$ distribution and azimuthal asymmetry, we present two sets of numerical results: one incorporating the $\text{Sud}^{ i}(b_\perp^*, P_\perp)$ factor from initial state gluon radiation and the other omitting it.

\begin{figure}[t!]
\centering{
    \includegraphics[width=0.49\textwidth]{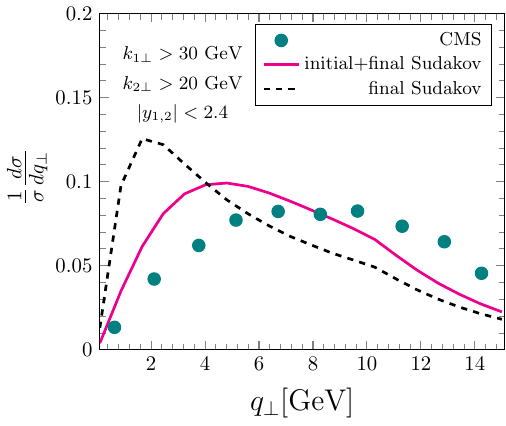}~
    \includegraphics[width=0.49\textwidth]{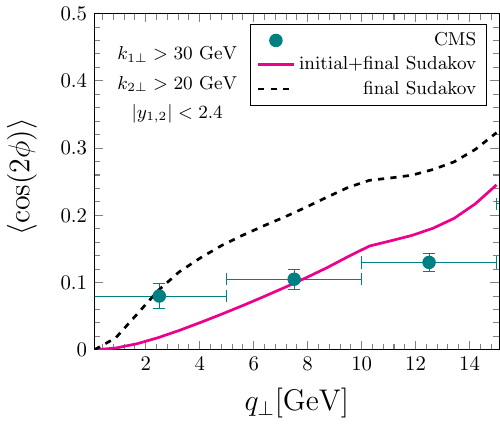}
    }
    \caption{ The normalized $q_\perp$ distribution computed at the LHC energy is  shown on the left panel. The right plot displays the $\cos (2\phi)$ azimuthal asymmetry as a function of $q_\perp$.  The experimental data points are taken from Ref.~\cite{CMS:2022lbi}.  Both observables are estimated with and without the Sudakov factor resulting from initial state radiations. }
    \label{fig:cms}
\end{figure}

The gluon dipole scattering amplitude ${\cal T}_g(x_{\mathbb{P}},r_\perp)$ is parametrized with the GBW model~\cite{Golec-Biernat:1998zce}, which reads
\begin{eqnarray}
 {\cal T}_g(x_{\mathbb{P}}, r_\perp) =1- \exp \left [  -\frac{1}{2} Q_p^2(x_{\mathbb{P}}) r_\perp^2  \right],
\end{eqnarray}
where 
\begin{equation}
Q_p^2(x_{\mathbb{P}})= Q_{0}^2\, (x_0/x_{\mathbb{P}})^\lambda, 
\end{equation}
with $x_0=3.04 \times 10^{-4}$, $\lambda=0.288$, and $Q_0^2=1\,$GeV$^2$. For a large nucleus target, the saturation scale is commonly  given by
\begin{equation}
Q_A^2(x_{\mathbb{P}})= s_0 A^{1/3} Q_p^2(x_{\mathbb{P}}),
\end{equation}
where $A$ is the atomic number of the nucleus and $s_0$ is the parameter representing the average centrality of the gluon dipole-nuclei collisions. In this work, we take $s_0=0.56$ to represent minimum bias collisions, which is the same parameter used in Refs.~\cite{Watanabe:2015tja}. 

Only events with highly asymmetrical configurations, where the cutoff for transverse momenta of two hard jets differ significantly, were selected in the CMS measurement~\cite{CMS:2022lbi}. In this case, it is more appropriate to choose the leading jet transverse momentum as the hard scale when performing the evolution, rather than $ P_\perp$. Additionally, we fix the various kinematic variables with the leading jet transverse momentum $k_{1\perp}$, for instance, the longitudinal momentum fractions $x_\gamma$, $x_g$ and $x_{q\bar q}$ are given as: 
\begin{align}
    x_{\gamma} = \frac{k_{1\perp}}{\sqrt{s}} \big( e^{y_1} + e^{y_2} \big),\quad {\rm and} \quad x_g = x_{q\bar q} = \frac{ k_{1\perp}}{\sqrt{s}} \big( e^{-y_1} + e^{-y_2} \big).
\end{align}  
The average value of the $\cos(2\phi)$ we compute numerically is defined as 
\begin{equation}
\langle \cos(2\phi) \rangle \equiv \frac{\int \md\mathcal{P}. \mathcal{S}.  \cos(2\phi) \frac{\md \sigma}{\md y_1 \md y_2 \md^2 \bm P_\perp \md^2 \bm q_\perp }  }{\int \md\mathcal{P}. \mathcal{S}. \frac{\md \sigma}{\md y_1 \md y_2 \md^2 \bm P_\perp \md^2 \bm q_\perp } }.
\end{equation}

The left panel of Fig.~\ref{fig:cms} displays the normalized $ q_\perp$ distribution of the di-jet system that is diffractively produced in UPCs for the CMS kinematics. Notably, incorporating the effect of initial state gluon radiation offers a more accurate representation of the CMS data~\cite{CMS:2022lbi}. Despite this improvement, a noticeable difference between our results and the CMS measurement remains. The azimuthal asymmetry is plotted as a function of $q_\perp$ on the right panel of Fig.~\ref{fig:cms}. Our result underestimates the observed asymmetry at low $ q_\perp$ and overshoots it at high $ q_\perp$.  It is worth noting that gluons inside a pomeron acquire finite transverse momentum through a Glauber gluon exchange between two $t$-channel gluons. This is why the asymmetry we computed without taking into account the initial state radiation effect is still suppressed at low $q_\perp$ compared to the result obtained in Ref.~\cite{Hatta:2021jcd}.

\begin{figure}[t!]
\centering{
    \includegraphics[width=0.49\textwidth]{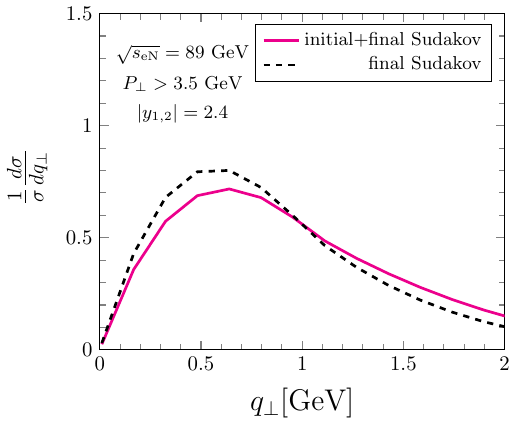}
    \includegraphics[width=0.49\textwidth]{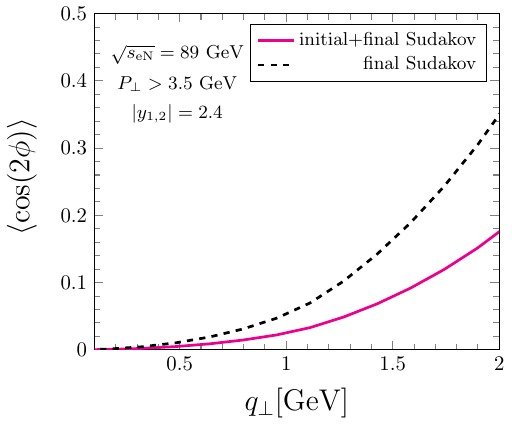}
    }
    \caption{The normalized ${ q}_\perp$ distribution(left panel) and the $ \cos(2\phi)  $  azimuthal asymmetry(right panel) in the diffractive photo-production of di-jet in eA collisions for the EIC kinematics.  }
    \label{fig:eic}
\end{figure}

\begin{figure}[t!]
\centering{
    \includegraphics[width=0.49\textwidth]{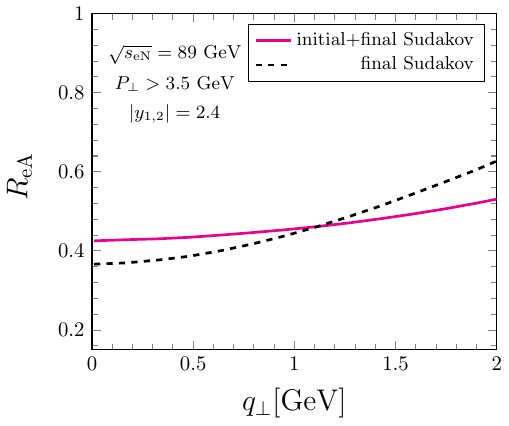}
    \includegraphics[width=0.49\textwidth]{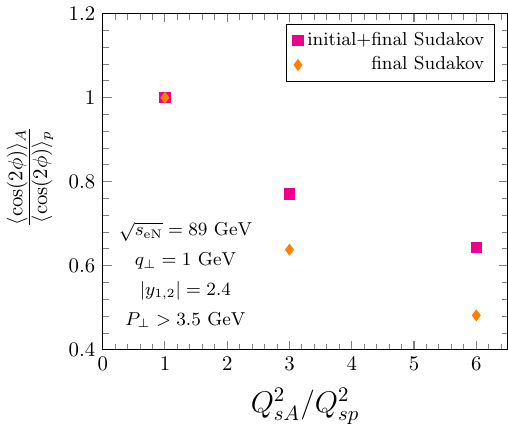}
    }
    \caption{The left panel displays the nuclear suppression factor for the normalized ${q}_\perp$ distribution. The nuclear suppression factor for the $\cos (2\phi)$ azimuthal asymmetry is plotted as the function of ${q}_\perp$ in the right panel.  Both ratios are computed at EIC energy.  }
    \label{fig:eic_ReA}
\end{figure}

We extend our analysis to include predictions for semi-inclusive diffractive photo-production of di-jets in $eA$ collisions within the EIC kinematics. The numerical results for both the normalized transverse momentum ($q_\perp$) distribution and the azimuthal asymmetry are illustrated in Fig.~\ref{fig:eic}. Moreover, we analyze identical observables for $ep$ collisions at the EIC, noting the enhanced saturation effects for a larger nucleus target. To quantitatively assess the saturation effect, we define the nuclear modification factor $R_{eA}$ as
\begin{equation}
    R_{eA} \equiv \frac{1}{A} \frac{\md\sigma_{eA}}{\md {\cal P.S. }}/\frac{\md\sigma_{ep}}{\md {\cal P.S. }},
\end{equation}
where $\md\mathcal{P.S.}$ denotes the phase space. For the evaluation of the pomeron gluon TMD, the transverse area of the heavy nucleus $Au$ is assigned a value of $S_{\perp}=  1830\,$mb, while the transverse area of the proton is assigned a value of  $S_{\perp}=  51 \,$mb.  In addition to $R_{eA}$, we introduce another physical quantity sensitive to the gluon saturation effect: the ratio of $\langle \cos(2\phi) \rangle$ for a heavy nucleus over that for a proton, denoted as  ${\langle \cos (2\Delta \phi) \rangle _A}/{\langle \cos (2\Delta \phi) \rangle _p}$. 

The $R_{eA}$ and ${\langle \cos (2\Delta \phi) \rangle _A}/{\langle \cos (2\Delta \phi) \rangle _p}$ computed with and without including initial state radiation effect are presented in Fig.~\ref{fig:eic_ReA}.  It is evident that the saturation effect leave its imprint on both observables. In particular, the suppression of ${\langle \cos (2\Delta \phi) \rangle _A}/{\langle \cos (2\Delta \phi) \rangle _p}$ amplifies as the saturation effect becomes stronger. This phenomenon opens a novel pathway to investigate the properties of highly dense gluonic matter in high-energy collisions. A related study on lepton-jet correlations in $eA$/$ep$ collisions is documented in Refs.~\cite{Tong:2022zwp, Tong:2023bus}.

\begin{figure}[t!]
\centering{
    \includegraphics[width=0.49\textwidth]{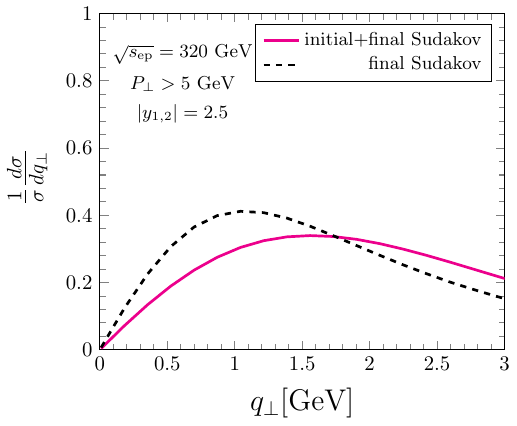}
    \includegraphics[width=0.49\textwidth]{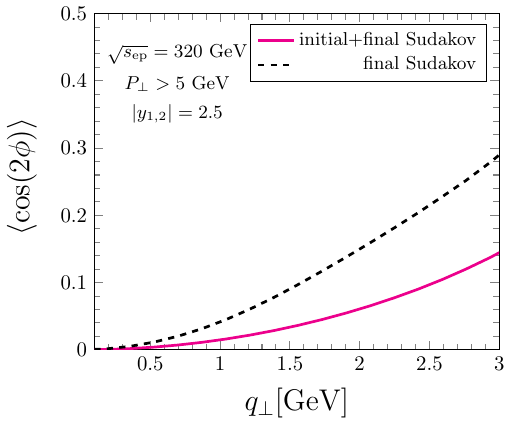}
    }
    \caption{The left panel displays the normalized $q_\perp$ distribution in diffractive photo-production of di-jet at HERA. 
 The $\cos(2\phi) $ azimuthal asymmetry is plotted as the function of $q_\perp$ at  HERA energy in the right panel. }
    \label{fig:hera}
\end{figure}

Furthermore, our predictions for the normalized $q_\perp$ distribution and the $\langle \cos(2\phi) \rangle$ within the HERA kinematic domain are displayed in Fig.~\ref{fig:hera}. It would be interesting to test our results against the EIC and HERA measurements in the future.

\section{Conclusion}\label{sec:conclusion}

We revisit the azimuthal angular asymmetry in diffractive di-jet production, inspired by the recent developments advanced by Iancu, Mueller, and Triantafyllopoulos. Their research has shown that semi-inclusive diffractive di-jet production dominates over exclusive di-jet production in photon-initiated scattering processes. This phenomenon is primarily attributed to tri-jet events characterized by an asymmetric setup, where a hard quark-antiquark dijet is accompanied by a semi-hard gluon jet that has been integrated out. Moreover, the production of color octet hard quark-anti-quark dijets at the Born level expands the color space, enabling the emission of soft gluons in the initial state. This mechanism significantly influences the total transverse momentum $q_\perp$ distribution of the dijet.

We have performed an all-order resummation for both initial and final state radiations, following the standard procedure. The pomeron gluon TMD, computed in the CGC formalism, is used as the input at the initial scale when performing the TMD evolution. The $ q_\perp$ distribution is found to be significantly broadened due to the effect of initial state radiation. This is because the initial state radiation effect has a leading double logarithm enhancement, whereas the leading contribution from final state radiation is the single logarithm term. We further investigated the impact of initial state radiation effects on the $\cos (2\phi)$ azimuthal asymmetry induced by the soft gluon emission from the hard jets. Unlike final state radiations, the transverse momentum distribution of soft gluons radiated from the incoming gluons exhibits axisymmetry. As a consequence, initial state radiation effects naturally lead to the suppression of the azimuthal asymmetry, which has been explicitly confirmed by our numerical calculations. Although our results quantitatively capture the overall trends in the $q_\perp$ distribution and the asymmetry observed by the CMS Collaboration, a sizable discrepancy between the experimental data and theoretical calculations remains. This might hint that the underlying mechanism behind diffractive di-jet production is not yet fully understood. To achieve a quantitative description of the CMS data, more theoretical efforts along this direction have to be made. We also made predictions for the same observables at both EIC and HERA energies. As a byproduct, we demonstrate that the $\cos (2\phi)$ asymmetry in diffractive di-jet production in $ep$/$eA$ collisions at EIC can serve as a sensitive probe of the saturation effect. In summary, as the normalized ${q}_\perp$ distribution and the azimuthal asymmetry are free of the uncertainties associated with the overall normalization of the cross-section, the study of these observables could provide us a unique opportunity to explore the production mechanism of hard dijet in diffractive processes.

\section*{Acknowledgements}
We thank Shu-Yi Wei for the valuable discussions.
D.Y.S.~is supported by the National Science Foundations of China under Grant No.~12275052 and No.~12147101 and the Shanghai Natural Science Foundation under Grant No. 21ZR1406100. J. Zhou has been supported by the National  Science Foundations of China under Grant No.\ 12175118 and the National Science Foundation under Contract No.~PHY-1516088. Y. Zhou has been supported by the Natural  Science Foundation of Shandong Province under Grant No. ZR2020MA098. C. Zhang has been supported by the National Science Foundations of China under Grant No.\ 12147125.  Y. Shi is supported by the China Postdoctoral Science Foundation under Grant No. 2022M720082. 

\appendix

\section{Soft function}\label{app:softfun}

In this appendix, we detail the calculation of the one-loop soft function using the narrow cone approximation, characterized by $R \ll 1$. 

The NLO soft function has be defined in Eq. \eqref{eq:nlo_softfun}. In order to simplify the calculation, we perform the phase space integration under the narrow cone limit, $R \ll 1$. The contribution outside the jet region is redefined as
\begin{align}
    \theta(\Delta R_i-R) = 1 - \theta(R-\Delta R_i),
\end{align}
with the first term indicating that soft radiation is independent of the jet definition. For $R \ll 1$, the remaining contribution describes the configuration of soft radiations near the jet boundary, often termed the collinear-soft function~\cite{Becher:2015hka,Buffing:2018ggv,Chien:2019gyf} within SCET framework. Explicitly, the NLO soft function becomes
\begin{align}
    S^{\rm NLO}(\bm b_\perp,R,\mu) = \frac{C_A}{2} s_{gq} + \frac{C_A}{2} s_{g\bar q} + \left(C_F-\frac{C_A}{2}\right) s_{q\bar q} + C_F c_q + C_F c_{\bar q},
\end{align}
where $s_{ij}$ denotes the contribution independent of the jet radius parameter $R$, and $c_{q(\bar q)}$ are the collinear soft functions, defined as follows:
\begin{align}
    s_{ij} = & \frac{\alpha_s \mu^{2 \epsilon} \pi^\epsilon e^{\gamma_E \epsilon}}{\pi^2} \int \md^d k \, \delta\!\left(k^2\right) \theta\!\left(k^0\right) \frac{n_i \cdot n_j}{n_i \cdot k \, k \cdot n_j}\left(\frac{\nu}{2 k^0}\right)^\eta e^{i \bm{k}_{\perp} \cdot \bm{b}_\perp}, \\
    c_i = & \frac{\alpha_s \mu^{2 \epsilon} \pi^\epsilon e^{\gamma_E \epsilon}}{\pi^2} \int \md^d k \,\delta\!\left(k^2\right) \theta\!\left(k^0\right) \frac{n_i \cdot \bar n_i}{n_i \cdot k \, k \cdot \bar n_i} \theta(\Delta R_i-R)  e^{-i \bar n_i \cdot k \, n_i \cdot b_\perp/2}.
\end{align}
After integrating, one obtains~\cite{Kang:2020xez,delCastillo:2020omr,Kang:2021ffh,delCastillo:2021znl}:
\begin{align}
    s_{g q}(\bm b_\perp,\mu) & = 
        \frac{\alpha_s}{2 \pi}\bigg[ \left(-\frac{2}{\eta}+\ln \frac{\mu^2}{\nu^2}+2 y_q+2 \ln \left(-2 i \cos \phi_{b}\right)\right)\left(\frac{1}{\epsilon}+\ln \frac{\mu^2b_\perp^2}{b_0^2}\right) +\frac{2}{\epsilon^2} +\frac{1}{\epsilon} \ln \frac{\mu^2b_\perp^2}{b_0^2} \notag \\
        & ~~~ -\frac{\pi^2}{6}\bigg], \\
    s_{q\bar q}(\bm b_\perp,\mu) & =\frac{\alpha_s}{2 \pi}\left[\frac{2}{\epsilon^2}+\frac{2}{\epsilon} \ln \frac{\mu^2 b_{\perp}^2}{b_0^2 A_\perp}+ \ln ^2 \frac{\mu^2 b_{\perp}^2}{b_0^2 A_\perp}+\frac{\pi^2}{2}-2 \ln A_\perp \ln \left(1-A_\perp\right)-2 \operatorname{Li}_2\left(A_\perp\right)\right], \notag 
\end{align}
with $s_{g\bar q}=s_{g q}|_{y_q\to y_{\bar q},\,\phi_b\to \phi_b + \pi}$, $b_0=2e^{-\gamma_E}$ and  $A_\perp=M^2 /\left(4 P_{\perp}^2 \cos ^2 \phi_{b}\right)$. Besides, the NLO collinear-soft function reads
\begin{align}
    c_q(\bm b_\perp,R,\mu)= & -\frac{\alpha_s}{2\pi} C_F\bigg[\frac{1}{\epsilon^2}+ \frac{1}{\epsilon}\ln \frac{\mu^2b_\perp^2}{b_0^2 R^2}+\frac{1}{2}\ln^2\frac{\mu^2b_\perp^2}{b_0^2 R^2}+ \frac{\pi^2}{4} +2\ln^2(-2i\cos\phi_{b}) \notag \\
    & +2\ln(-2i\cos\phi_{b})\left(\frac{1}{\epsilon}+\ln\frac{\mu^2 b_\perp^2}{b_0^2 R^2}\right)\bigg],
\end{align}
with $c_{\bar q} = c_q|_{\phi_b\to \phi_b+\pi}$. Then one can apply the standard RG and the Collins-11 techniques to renormalize the $\epsilon$ and $\eta$ poles, respectively~\cite{Boussarie:2023izj}. Finally, we obtain the one-loop soft factor, and the azimuthal angle dependent terms are given in Eq. \eqref{eq:s_NLO_phidep}. 

\bibliographystyle{JHEP}
\bibliography{ref}

\end{document}